\documentclass[a4paper,11pt]{article}
\pdfoutput=1 
\usepackage{jheppub} 
\usepackage[T1]{fontenc} 
\usepackage{mathrsfs}
\newcommand{\AdS} {{AdS}}
\usepackage{shuffle}
\usepackage{musicography}
\usepackage{empheq}
\usepackage{mathrsfs}
\usepackage{bm}
\usepackage{verbatim}
 \usepackage{makecell}
 \usepackage{float}
 \usepackage{hyperref}
 \usepackage[makeroom]{cancel}

\newcommand{\be}{\begin{equation}}
\newcommand{\ee}{\end{equation}}
\newcommand{\bpm}{\begin{pmatrix}}
\newcommand{\epm}{\end{pmatrix}}

\newcommand{\PBK}[1]{\ensuremath{\begin{pmatrix}#1\end{pmatrix}}}

\newcommand{\EV}[1]{\langle #1 \rangle}
\newcommand{\beqn}{\begin{eqnarray}}
\newcommand{\eeqn}{\end{eqnarray}}

\usepackage{slashed}

\usepackage{pgffor}
\foreach \x in {A,...,Z}{
  \expandafter\xdef\csname b\x\endcsname{\noexpand\mathbb{\x}}
}
\foreach \x in {A,...,Z}{
  \expandafter\xdef\csname c\x\endcsname{\noexpand\mathcal{\x}}
}
\foreach \x in {A,...,Z}{
  \expandafter\xdef\csname s\x\endcsname{\noexpand\mathscr{\x}}
}
\foreach \x in {A,...,Z}{
  \expandafter\xdef\csname sf\x\endcsname{\noexpand\mathsf{\x}}
}
\foreach \x in {a,...,z}{
  \expandafter\xdef\csname sf\x\endcsname{\noexpand\mathsf{\x}}
}
\foreach \x in {A,...,Z}{
  \expandafter\xdef\csname  fk\x\endcsname{\noexpand\frak{\x}}
}
\foreach \x in {a,...,z}{
  \expandafter\xdef\csname  fk\x\endcsname{\noexpand\frak{\x}}
}

 \newcommand{\bv}{ \begin{verbatim}}

\newcommand{\ii}{\mathrm{i}}
\newcommand{\dd}{\mathrm{d}}

\title{ $AdS_3 \times S^3$ Virasoro-Shapiro amplitude with KK modes}

\abstract{We study the first curvature correction to the string amplitude of four Kaluza–Klein (KK) modes on $AdS_3 \times S^3 \times M_4$, with $M_4=K3$ or $T^4$, in type IIB string theory, which is holographically dual to the four-point correlator $\langle \mathcal{O}_{p_1} \mathcal{O}_{p_2} \mathcal{O}_{p_3} \mathcal{O}_{p_4} \rangle$ of certain half-BPS operators in the boundary D1-D5 CFT. The result takes the form of an integral over the Riemann sphere, analogous to the flat-space Virasoro–Shapiro amplitude, but with insertions of single-valued multiple polylogarithms of weight three.
Our results are obtained in two steps. First, we derive the $AdS_3 \times S^3$ Virasoro–Shapiro amplitude in the special case $\langle \mathcal{O}_{p} \mathcal{O}_{p} \mathcal{O}_{1} \mathcal{O}_{1} \rangle$, by matching the CFT block expansion with an ansatz based on single-valued multiple polylogarithms. We then employ the $\AdS \times S$ Mellin formalism to generalize the result to the  general case of four arbitrary KK modes  $\langle \mathcal{O}_{p_1} \mathcal{O}_{p_2} \mathcal{O}_{p_3} \mathcal{O}_{p_4} \rangle$.
Our analysis yields an infinite set of results for operator anomalous dimensions and OPE data in D1-D5 CFT at strong coupling. In particular,   the resulting scaling dimensions of certain operators are shown to be consistent with classical string theory computations.
}

\author[a,b,c]{Hongliang Jiang}
\author[c]{and De-liang Zhong}
\affiliation[a]{Center for Mathematics and Interdisciplinary Sciences, Fudan University, Shanghai 200433, China}
\affiliation[b]{Shanghai Institute for Mathematics and Interdisciplinary Sciences,  657   
Songhu Road, Shanghai 200433, China}
\affiliation[c]{Abdus Salam Centre for Theoretical Physics, Imperial College London, London SW7 2AZ, UK}
\emailAdd{jianghongliang@fudan.edu.cn, d.zhong23@imperial.ac.uk}

\begin{document}
\maketitle

\section{Introduction}

The AdS/CFT correspondence has become an indispensable tool for studying quantum gravity, offering a quantitative, precise, and non-perturbative framework to probe the fundamental nature of gravity. Establishing a rigorous proof of the AdS/CFT correspondence remains a central challenge in theoretical physics. Among various cases, one of the most promising candidates is the duality between $\mathcal{N}=4$ Super Yang--Mills (SYM) theory and type IIB string theory on $\AdS_5 \times S^5$, due to the maximal supersymmetry it preserves. Over the past few decades, significant progress has been made in understanding $\mathcal{N}=4$ SYM through various approaches, including localization \cite{Pestun:2016zxk} and integrability \cite{Beisert:2010jr,Gromov:2013pga}. However, directly computing physical observables from the worldsheet formulation of type IIB string theory on $\AdS_5 \times S^5$ remains a formidable challenge, primarily due to the difficulty of handling Ramond--Ramond (RR) flux backgrounds.

 Nevertheless, remarkable progress has been made in the  past few years by providing a worldsheet description of graviton scattering in AdS space. Instead of deriving the worldsheet formulation directly from string theory,     an ansatz  was proposed for the worldsheet integral involving single-valued multiple polylogarithms (SVMPLs), which was fully fixed by comparison with the flat space Virasoro--Shapiro amplitude and consistency with CFT expectations \cite{Alday:2022uxp,Alday:2022xwz,Alday:2023jdk,Alday:2023mvu}. The resulting expression is therefore referred to as the AdS Virasoro-Shapiro (VS) amplitude.  This approach circumvents the difficulties associated with Ramond--Ramond fluxes and is powerful enough to determine the first non-trivial curvature correction in various AdS backgrounds \cite{Alday:2024yax,Alday:2024ksp,Chester:2024wnb,Chester:2024esn}.\footnote{The second-order curvature correction can also be determined with the help of supersymmetric localization and integrability. More precisely, integrability provides the second-order correction to the leading Regge trajectory operator, while localization imposes integrated correlator constraints \cite{Binder:2019jwn, Chester:2020dja}. 
}

 The aim of this paper is to generalize this method to the case of $\AdS_3$ supported by RR fluxes, including Kaluza--Klein (KK) modes on $S^3$, building upon previous works \cite{Alday:2023jdk,Chester:2024wnb,Fardelli:2023fyq,Wang:2025pjo}. 
More precisely, we   consider   type IIB string theory on $\AdS_3\times S^3\times M_4$ for $M_4=K3$ or $T^4$, which can be obtained from the near horizon limit of D1-D5 system  by wrapping $Q_1$ D1
branes along a non-compact direction, and $Q_5$ D5  branes along  $M_4$ and  
the non-compact direction shared with the D1 branes.
The holographic dual   is given by a 2d $\mathcal  N = (4, 4)$  superconformal field theory (SCFT)
  arising at the low energy bound states of D1-D5 branes and it is hence   called  D1-D5  CFT, whose direct field theory realization remains unclear except at certain special points on the moduli space.\footnote{
 For type IIB string theory on $\AdS_3 \times S^3 \times M_4$ with minimal NS-NS flux, the holographic dual is given by the symmetric orbifold CFT $\mathrm{Sym}^N(M_4)$, which has been shown from different perspectives \cite{Eberhardt:2018ouy,Eberhardt:2019ywk}. This has been made possible by the existence of a worldsheet description of string theory on $\AdS_3$ with NS--NS flux in terms of the WZW model.
Although RR and NS--NS fluxes are related by S-duality in string theory, a useful and satisfactory worldsheet formulation for $\AdS_3$ backgrounds supported by RR flux is still lacking—a challenge closely analogous to the $\AdS_5$ case. 
}   
The curvature of AdS and the couplings in CFT are related as: 
 $ 
\frac{R_{\text{AdS}}^2}{\alpha^\prime}=gN \equiv\sqrt\lambda$,
where $R_{\text{AdS}}$ is the $  \AdS_3$ radius,   $g$ is the string coupling in 6d,   and $N=\sqrt{Q_1 Q_5}$. 
See \cite{Aharony:2024fid,David:2002wn} for more details about the D1-D5   system.

The tree-level scattering amplitude of dilatons in $\AdS_3 \times S^3 \times M_4$ is dual to the correlation function of a certain   half-BPS multiplet in the two-dimensional CFT, in the large-$N$ planar limit and valid to all orders in $\lambda$. This setup was studied in~\cite{Chester:2024wnb}, where extensive CFT data were derived by combining a worldsheet ansatz involving single-valued multiple polylogarithms   with CFT consistency conditions. 

In this paper, we generalize those results by studying  the  AdS Virasoro-Shapiro  amplitude for the  half-BPS operators corresponding to Kaluza–Klein (KK) modes on $S^3$. The corresponding half-BPS KK operator is denoted $\cO_p$ and has scaling dimension $p$, where $p = 1, 2, \cdots$. The four-point correlators of interest are then given by $\EV{\cO_{p_1} \cO_{p_2} \cO_{p_3} \cO_{p_4}}$, or more simply $\EV{p_1p_2p_3p_4}$, which generalize the simplest  case $\EV{1111}$ studied in~\cite{Chester:2024wnb}.
Our derivation proceeds in two steps. First, we derive the $\AdS_3 \times S^3$ Virasoro–Shapiro amplitude for the special case $\EV{pp11}$, by matching the CFT block expansion with a worldsheet ansatz based on single-valued multiple polylogarithms. Then, following \cite{Wang:2025pjo}, we use the $\AdS \times S$ Mellin formalism to generalize this result to the case of $\EV{p_1p_2p_3p_4}$, describing the correlators of four arbitrary KK modes.

The rest of the paper is organized as follows. In Section~\ref{4ptfcn}, we introduce our setup from the $\AdS_3$ string theory perspective and review the kinematic properties of four-point functions in two-dimensional SCFTs, both in position space and Mellin space, as well  their flat-space limits. In Section~\ref{wspp11}, we derive the Virasoro–Shapiro amplitude for $\EV{pp11}$, providing a worldsheet description involving    SVMPLs. In Section~\ref{wsp1234}, we employ the $\AdS \times S$ Mellin formalism to compute the Virasoro–Shapiro amplitude for general correlators $\EV{p_1p_2p_3p_4}$. In Section~\ref{concl}, we summarize our main results and outline possible  future directions.
Finally, we also relegate some technical details to the appendices.

\section{Four-point  amplitudes for $\AdS_3$ in general } \label{4ptfcn}

\subsection{$\AdS_3$ KK spectrum}
We consider type IIB string theory on \( \AdS_3 \times S^3 \times M_4 \), where \( M_4 = K3 \). The low-energy limit is described by six-dimensional \( \mathcal{N} = (2,0) \) supergravity on \( \AdS_3 \times S^3 \), coupled to \( h_{1,1} + 1 \) tensor multiplets, with \( h_{1,1}(K3) = 20 \).\footnote{One can also consider the case \( M_4 = T^4 \), in which the low-energy theory is six-dimensional \( \mathcal{N} = (2,2) \) supergravity coupled to \( h_{1,1} + 1 \) tensor multiplets, where \( h_{1,1}(T^4) = 4 \), along with additional fermionic multiplets that do not affect tree-level bosonic correlators. For concreteness, we focus on the \( K3 \) case in the main text, while keeping \( h_{1,1} \) general.}
Further compactifying on $S^3$ gives rise to various towers of KK modes in $\AdS_3$. See \cite{Rastelli:2019gtj} for a summary.
 
The underlying symmetry is given by $PSU(1,1|2)_L\times PSU(1,1|2)_R$ for the left and right moving sectors.\footnote{The symmetry algebra is further enhanced to the small $\mathcal N=4$  super-Virasoro algebra for $M_4=K3$. The enhancement does not play a role in this paper, and thus will not be considered.}
The bosonic subgroup is $ SL(2,\mathbb R)_L\times SU(2)_L\times SL(2,\mathbb R)_R\times SU(2)_R $, which matches the isometry group $SO(2,2)\times SO(4) $ of the $\AdS_3 \times S^3$ background.  
Each operator is labeled by the quantum numbers $h,j,\bar h, \bar j$, corresponding to the Cartan generators of the bosonic subgroup. The scaling dimension $\Delta$ and spin $\ell$ of the operator are then given by $\Delta=h+\bar h, \ell=h-\bar h$.

In this paper, we are interested in the $1/2$-BPS operators coupling to the tensor multiplets, whose superconformal primaries are denoted as $s_p^I$, where $p=1, 2,\cdots$ labels the KK modes under the reduction on $S^3$, and $I=1, \cdots, h_{1,1}+1$ labels the vector representation of the flavor symmetry group $SO(h_{1,1}+1)$. The quantum numbers are given by $h=j=\bar h =\bar j=p/2$.

 The flavor symmetry $SO(h_{1,1}+1)$ is valid in the supergravity limit and is respected by the cubic coupling. However, in full string theory, this symmetry is broken down to $SO(h_{1,1})$. See e.g. \cite{Taylor:2007hs}.   We will thus focus on the unique singlet under the $SO(h_{1,1})$, and denote it as $\cO_p$. For $p=1$, the multiplet $\cO_1$ includes a marginal operator that couples to the string coupling, which is dual to the dilaton of string theory. In the planar limit that we consider, this   multiplet   does not care about the compact spacetime  $M_4$   and is invariant under   certain  duality symmetries, including T-duality. For higher $p>1$, $\cO_p$ can be regarded as the cousins    of $\cO_1$ with higher KK modes under the reduction on $S^3$. The goal of this paper is to study the correlation function of four arbitrary $\cO_p$'s beyond the supergravity limit.  

\subsection{Superconformal kinematics in position space}

The operator $\cO_p$ transforms as spin-$(p/2,p/2)$ representation under the $SU(2)_L \times SU(2)_R$ R-symmetry, and   carries $p$ symmetric vector indices under $SO(4)$. To simplify notation, we introduce an index-free form
\be
\cO_p(\sfx;\sfy) \coloneq \cO_p^{\mu_1 \cdots \mu_p} (\sfx) \sfy_{\mu_1}  \cdots \sfy_{\mu_p}~,
\ee
where $\sfy _\mu$ is an $SO(4)$ null vector satisfying $\sfy_\mu \sfy^\mu=0$. Using the isomorphism $SO(4)\simeq SU(2) \times SU(2)$, we can parametrize $\sfy^\mu=\sigma_{\alpha\dot\alpha}^\mu y^\alpha \bar y ^{\dot \alpha}$ and label $\cO$ in spinor representation 
 \be
 \cO_p(\sfx; \sfy )=  \cO_p(\sfx; y,\bar y ) =\cO_p^{\alpha_1\cdots\alpha_p, \dot\alpha_1\cdots \dot\alpha_p}
 y_{\alpha_1}\cdots y_{\alpha_p}\;\bar y_{\dot\alpha_1}\cdots \bar y_{\dot\alpha_p}~.
 \ee
 Similarly, instead  of using the real Euclidean coordinate $\sfx^1, \sfx^2$ to label the position of operators in the dual 2d CFT, we introduce complex coordinates $z=\sfx^1+ \ii\, \sfx^2,\bar z=\sfx^1-\ii\, \sfx^2$. The Lorentz-invariant distance between two points is then given by
 \be
 \sfx_{ij}^2=(\sfx^1_i-\sfx_j^1)^2+(\sfx^2_i-\sfx_j^2)^2=z_{ij}\bar z_{ij}~, \quad \text{with} \quad z_{ij}=z_i -z_j~, \;\bar z_{ij}=\bar z_i -\bar z_j\,~.
 \ee  
For the R-symmetry polarization vectors, we define analogously $\sfy_{ij}=\sfy_i^\mu \sfy _{ j\,  \mu}= y_{ij}  \bar y_{ij}$, where $y_{ij}=\epsilon_{\alpha\beta} y_i^\alpha y_j^\beta, \bar y_{ij}=\epsilon_{\dot\alpha\dot\beta} y_i^{\dot\alpha} y_j^ {\dot \beta}$.\footnote{By rescaling, one can choose  $y_i  = \PBK{1 \\  \mathrm y_i} $, then $y_{ij}=\mathrm y_i -\mathrm y_j$.}

We can construct conformal and R-symmetry cross ratios from the kinematic building blocks introduced above. There are two conformal cross ratios, $U$ and $V$, and two R-symmetry cross ratios, $\sigma$ and $\tau$, defined as
\be
U=\frac{\sfx_{12}^2 \sfx_{34}^2}{\sfx_{13}^2 \sfx_{24}^2}=z \bar{z}~, 
\qquad V=\frac{\sfx_{14}^2 \sfx_{23}^2}{\sfx_{13}^2 \sfx_{24}^2}=(1-z)(1-\bar{z})~ , 
\ee
and 
\be
 \sigma =\frac{\sfy_{13} \sfy_{24}}{\sfy_{12} \sfy_{34}}=\alpha \bar{\alpha}~, 
\qquad
 \tau =\frac{\sfy_{14} \sfy_{23}}{\sfy_{12} \sfy_{34}}=(1-\alpha)(1-\bar{\alpha})~ . 
 \ee
In terms of spinor variables, the cross ratios are equivalently given by:
\be
z=\frac{z_{12} z_{34}}{z_{13} z_{24}}~, \quad \bar{z}=\frac{\bar{z}_{12} \bar{z}_{34}}{\bar{z}_{13} \bar{z}_{24}}~, \quad \alpha=\frac{1}{y}= \frac{y_{13} y_{24}}{y_{12} y_{34}}~, \quad \bar{\alpha}=\frac{1}{\bar y}=\frac{\bar{y}_{13} \bar{y}_{24}}{\bar{y}_{12} \bar{y}_{34}}~ .
\ee

It is also convenient to introduce the following alternative R-symmetry cross ratios:
\beqn
\widetilde U &=&\frac {\sfy_{12} \sfy_{34}}{\sfy_{13} \sfy_{24}}=y \bar{y}=\frac{1}{\alpha\bar\alpha}=\frac{1}{\sigma}~, \qquad
\\
\widetilde V &=&\frac {\sfy_{14} \sfy_{23}}{\sfy_{13} \sfy_{24}}=(1-y)(1- \bar{y})=\frac{ (\alpha-1)(\bar \alpha-1)}{\alpha\bar\alpha}
=\frac{ (1-\alpha )(1-\bar \alpha )}{\alpha\bar\alpha}=\tau/\sigma~.
\eeqn

Finally, in superconformal kinematics, it is useful to define the combinations:
\be
g_{ij}=\frac {y_{ij}}{z_{ij}}~, \qquad \bar g_{ij}=\frac{\bar y_{ij}}{\bar z_{ij}} ~, \qquad
 \bm g_{ij}=g_{ij} \bar g_{ij}~.
\ee

Without loss of generality, we will assume $p_1\ge p_2 \ge p_3\ge p_4$, and study the four-point function: 
\be
\EV{ \cO_{p_1}(\sfx_1; \sfy_1 )  \cO_{p_2}(\sfx_2; \sfy_2 )
  \cO_{p_3}(\sfx_3; \sfy_3 )   \cO_{p_4}(\sfx_4; \sfy_4 ) }
 = {\bf K }  \cG_{p_1p_2p_3p_4}(z, \bar z ; \alpha, \bar \alpha)~,
\ee
which will be abbreviated as $\EV{p_1p_2p_3p_4}$. The prefactor $\mathbf{K}$ captures the kinematic dependence and is given by
\beqn
{\bf K}&=&\bm g_{12}^{\frac{p_1+p_2-p_3-p_4}{2}}\bm g_{13}^{\frac{p_1+p_3-p_2-p_4}{2}}\bm g_{14}^{p_4}\bm \;\bm g_{23}^{\frac{p_2+p_3+p_4-p_1}{2}}
\Big(\frac{\bm g_{12}\bm g_{34}} {\bm g_{14}\bm g_{23}}\Big) ^L
\\&=&  \label{Kfactor}
\bm g_{12}^{\frac{p_1+p_2-p_3-p_4}{2}}\bm g_{13}^{\frac{p_1+p_3-p_2-p_4}{2}} \bm \;\bm g_{23}^{\frac{p_2+p_3-p_4-p_1}{2}}(\bm g_{14}\bm g_{23})^{p_4}
\Big(\frac{V}{U \tau } \Big) ^L~ .
\eeqn
Here we used the relations:
\be
 \frac{\bm g_{12}\bm g_{34}} {\bm g_{14}\bm g_{23} }=\frac{V\tilde U}{U\tilde V}=\frac{V/\sigma}{U \tau /\sigma}=\frac{V}{U \tau }~, 
 \qquad
 \frac{\bm g_{14}\bm g_{23}} {\bm g_{13}\bm g_{24} } 
 =\frac{\tilde V}{V} =\frac{\tau }{V\sigma}~.
\ee
 
The function $\cG$ satisfies the superconformal Ward identities and admits the decomposition \cite{Rastelli:2019gtj}:
\be
\cG=\cG_0 +(1-z\alpha) (1-\bar z \bar \alpha)  \cH(U,V; \sigma, \tau)~,
\ee
where $\cG_0$ is the protected part that obeys
\be
\cG_0(z,\bar z; \alpha, \bar\alpha=1/\bar z)=f(z,\alpha)~, \qquad
\cG_0(z,\bar z; \alpha=1/z, \bar\alpha )=f(\bar z,\bar \alpha) ~,
\ee
and survives in the free limit due to non-renormalization theorems.

The {reduced correlator} $\cH$ receives contributions from both protected and unprotected operators, coming respectively from the short and long multiplets of the SCFTs. In this work, we will focus on contributions from the unprotected long multiplets. Importantly, $\cH$ is a polynomial in $\alpha$ and $\bar \alpha$ of degree $L - 1$ \cite{Rastelli:2019gtj}, where the \textit{extremality} $L$ is defined as 
\be\label{extremL}
L \coloneq \frac{p_4-p_1}{2}+\frac12 \min(p_1+p_4,p_2+p_3)
=\begin{dcases} 
p_4 & \text{if}\ p_4-p_3\le p_2-p_1\\
\frac{p_2+p_3+p_4-p_1}{2} &\text{if}\ p_4-p_3\ge p_2-p_1 \\
\end{dcases}~ .
\ee

\subsection{Mellin amplitude and Borel transformation   }

The Mellin amplitude for the reduced correlator is defined by the following integral transform \cite{Rastelli:2019gtj,Behan:2024srh}  
\beqn \label{Mellin}
\hspace{-8ex} \cH(U,V,\sigma, \tau )&=&\int_{-\ii\infty}^ {\ii \infty}\frac{\dd s\, \dd t}{(4\pi \ii)^2}\, U^{\frac{s-p_3-p_4}{2}+L}\, V^{\frac{t+p_4-p_1}{2}-L }\,  M(s,t,\sigma, \tau )\nonumber
\\&&
\times \Gamma\Big(\frac{p_1+p_2-s}{2} \Big)  \Gamma\Big(\frac{p_3+p_4-s}{2} \Big) 
\Gamma\Big(\frac{p_1+p_4-t}{2} \Big)  \Gamma\Big(\frac{p_2+p_3-t}{2} \Big)
\\ &&
\times
 \Gamma\Big(\frac{p_1+p_3-\tilde u}{2} \Big)  \Gamma\Big(\frac{p_2+p_4-\tilde u}{2} \Big)~ . 
 \nonumber
 \eeqn
Here, the variables $s, t, \tilde{u}$ are analogous to the standard Mandelstam variables in flat-space four-point amplitudes, and they satisfy $s+t+\tilde u =p_1+p_2 + p_3 +p_4-2$.

To obtain the curvature expansion near flat space, one has to perform an additional Borel transformation to the Mellin amplitude, which defines the AdS Virasoro-Shapiro amplitude $A$ as  \cite{Alday:2024ksp,Penedones:2010ue} 
\be\label{Aborel}
 A(S,T,\sigma, \tau )=\lambda^{\frac12}\Gamma(\Sigma) \int\frac{\dd\alpha}{2\pi \ii}\frac{e^\alpha}{\alpha^ {\Sigma+1}} 
 M\Big(\frac{2\sqrt\lambda S}{\alpha}+\frac{2\Sigma -2}{3},\frac{2\sqrt\lambda T}{\alpha}+\frac{2\Sigma -2}{3},\sigma, \tau\Big)~,
 \ee
 where we define
 \be
 \Sigma=\frac12(p_1+p_2+p_3+p_4)~ .
 \ee

Physically, the Borel transformation provides a  resurgent resummation of the asymptotic strong coupling expansion of $\mathcal{M}$, yielding a well-defined amplitude that reorganizes the curvature corrections in power series of $\lambda^{-1/2}$ in the flat-space limit.

In the strong coupling or large AdS radius limit, $A$ reduces to the flat space Virasoro-Shapiro amplitude:\footnote{Note that the flat-space limit of the AdS Virasoro--Shapiro amplitude is universal for all correlators   \(\langle p_1 p_2 p_3 p_4 \rangle\), since all KK modes become massless in that limit. In the simplest case of \(\langle 1 1 1 1 \rangle\), the amplitude is also related to the flat-space scattering amplitude of four dilatons in type IIB string theory, as computed for example in~\cite{Okuda:2010ym}. This result agrees with~\eqref{flatVS} up to an overall factor of \(S^2 + T^2 + U^2\). The discrepancy arises because the dilaton couples to a dimension-two marginal operator in the multiplet, whereas in our case we always study superconformal primaries, which for \(\langle 1 1 1 1 \rangle\) have scaling dimension one.
 }
\be \label{flatVS}
A^{(0)}(S,T)=-\left(S^2+T^2+U^2\right)\frac{\Gamma (-S) \Gamma (-T) \Gamma (-U)  }{4 \Gamma (S+1) \Gamma (T+1) \Gamma (U+1)}~, \quad S+T+U=0~ .
\ee
  
\section{Virasoro-Shapiro amplitude for $\EV{pp11}$} \label{wspp11}

 In this section, we derive the Virasoro–Shapiro amplitude for $\EV{pp11}$ by employing a worldsheet ansatz involving  SVMPL, and by imposing consistency with the conformal block expansion in CFT, the flat space limit, and the supergravity limit.
For the correlator $\EV{pp11}$ considered here, the extremality parameter is $L = 1$ (see \eqref{extremL}). As a result, the reduced correlator $\cH$, the Mellin amplitude $M$, and the Virasoro–Shapiro amplitude $A$ are all \emph{independent} of the R-symmetry cross ratios $\sigma$ and $\tau$.

 \subsection{Supergravity limit}

 Let us first understand the supergravity limit of the AdS VS amplitude, which has been studied in Mellin space.
 
 Given the Mellin amplitude, we can consider its  Borel transformation  \eqref{Aborel} which, in the current case, reduces to 

 \be \label{Borel11pp}
 A(S,T )=\lambda^{\frac12} \Gamma(p+1)\int\frac{\dd\alpha}{2\pi \ii}\frac{e^\alpha}{\alpha^ { p+2}}\, M\left(\frac{2\sqrt\lambda S}{\alpha}+\frac{2p}{3},\frac{2\sqrt\lambda T}{\alpha}+\frac{2p}{3} \right)~.
 \ee
 
  It satisfies the crossing relation:
 \be
 A(S,T)=A(S,U) ~, \qquad \text{where} \qquad S+T+U=0~.
 \ee

The AdS amplitude can be expanded around flat space as
\be \label{eqn-pp11-Aexp}
\begin{aligned}
 A(S,T)&=A^{(0)}(S,T)+\frac{1}{\sqrt{\lambda}}A^{(1)}(S,T)+\cdots~,\\ 
 \end{aligned}
\ee
where $A^{(0)}(S,T)$ is the flat-space VS amplitude \eqref{flatVS}, and the rest of terms are curvature corrections.

At each order $A^{(k)}$ in \eqref{eqn-pp11-Aexp}, the leading term  corresponds to the supergravity contribution and is singular in the small momentum limit. It can be obtained  from the Mellin amplitude in the supergravity limit, which has been studied extensively \cite{Giusto:2019pxc}: 
 \be
M_{\text{SG}}(s,t)=-\frac{1}{s}-\frac{1}{t+1-p}-\frac{1}{\tilde u+1-p}~, 
\qquad \tilde u= 2p-s-t~ .
\ee 
Inserting it to \eqref{Borel11pp} yields
  \beqn
 A_{\text{SG}} 
& =& 
A_{\text{SG}}^{(0)} +\frac{1}{\sqrt\lambda}A_{\text{SG}}^{(1)} +\cdots \nonumber
\\ 
& =& 
\frac{S^2+T^2+U^2}{ 4 S T U }
+\frac{ p }{12 \sqrt{\lambda } }
\Big(    \frac{2p }{S^2}-
\frac{p-3}{T^2}
-
\frac{p-3}{U^2}
\Big)+\cO\left(\frac 1 \lambda \right)~ .
  \eeqn

As a result, we can write both  amplitudes as  $M=M_{\text{SG}}+M_{\text{string}}$ and $A=A_{\text{SG}}+A_{\text{string}}$, where the stringy non-supergravity parts are all polynomials in the Mandelstam-like variables. 

From the CFT perspective, 
the Mellin amplitude $M$ receives contributions from both the short and long multiplets, just like the reduced correlator $\cH$ which is related to $M$ via \eqref{Mellin}. 
  The protected operators completely determine the supergravity part of the amplitude, while   long multiplet leads to the stringy  corrections to the CFT data.  In the planar  or large-$N$ limit, those single-trace stringy operators  have  large anomalous dimension or twist of order $\cO(\lambda^{\frac{1}{4}})$. One goal in the rest of the section is to study the CFT data associated with these stringy operators. 
  
 \subsection{Poles from     block expansion}

To find those stringy corrections, we need to consider the block expansion in CFT. In Mellin space, we can expand the Mellin amplitude in terms of the Mellin blocks corresponding to $S$-channel and $T$-channel exchanged operators  
\be \label{eqn-MexpSTchannel}
M(s,t)=\sum_{\tau_s, \ell} C_s(\tau_s,\ell)\sum_{m=0}^\infty   \frac{\cQ_s(\tilde u; \tau_s, \ell, m)}{s-\tau_s-2m }
+\sum_{\tau_t, \ell} C_t(\tau_t,\ell)\sum_{m=0}^\infty   \frac{\cQ_t(\tilde u; \tau_t, \ell, m)}{t-\tau_t-2m }~,
\ee
where $\tilde u = 2p-s-t$. This formula can be derived from the contour deformation on the complex $s$ plane by picking up the $S$- and $T$-channel poles    with $\tilde u$ fixed.\footnote{Strictly speaking, the equality in Eq.~\eqref{eqn-MexpSTchannel} holds only under certain Regge boundedness conditions, which can be justified by appealing to the chaos bound. More generally, the left-hand side and right-hand side may differ by entire functions, such as polynomial terms in   $s$ and  $t$. However, since our analysis relies only on the behavior near the poles, the presence of such entire terms does not affect our derivation or results.  \label{ft6} }
Since the derivation closely parallels the $\AdS_5$ case studied in~\cite{Fardelli:2023fyq}, we omit the details here and refrain from repeating the steps of derivation.

In \eqref{eqn-MexpSTchannel}, $C_s,C_t$ denotes the product OPE coefficient of the exchanged operators,\footnote{
These are also exactly the coefficients appearing in the superconformal block expansion of the reduced correlator \cite{Aprile:2021mvq}. For example, in the $S$-channel, we have 
\begin{equation*}
\cH(U,V)=\sum_{\tau_s,\ell}C_s(\tau_s,\ell) U^{-1}g_{\tau_s+2,\ell}(U,V)
\end{equation*}
where 
 $  
g_{\tau,\ell}(U,V) = \frac{k_{\tau+2\ell}(z)\,k_{\tau}(\bar{z}) + k_{\tau+2\ell}(\bar{z})\,k_{\tau}(z)}{1 + \delta_{\ell,0}},
\, 
k_\beta(z) = z^{\beta/2} {}_2F_1\left(\beta/2, \beta/2, \beta; z\right).$ 
There is also a similar but a bit more complicated expression for the $T$-channel.

} and $\tau\equiv \Delta-\ell, \ell$ denote the twist  and the spin of the exchanged operators, respectively. The residue at each simple pole is given by the following Mack polynomials,
\be \label{cQs}
\begin{aligned}
 \cQ_s(w; \tau , \ell, m) & \coloneq \kappa_{\ell, m, \tau +2,d=2}^{\left( p+1, p+1,2,2\right)}Q_{\ell,m}^{0,0,\tau+2}(w-p-1)~ , \\
 \cQ_t(w; \tau , \ell, m) & \coloneq \kappa_{\ell, m, \tau +2,d=2}^{\left( p+1,  2,2,p+1\right)}Q_{\ell,m}^{p-1, 1-p,\tau+2}(w-2p )~,
\end{aligned}
\ee
where the precise definitions of the Mack polynomials and normalization factors are provided in Appendix~\ref{apd-mackpol}.

To proceed, we perform the Borel transformation~\eqref{Borel11pp} on the right-hand side of~\eqref{eqn-MexpSTchannel}. This yields the following expression:\footnote{Just like Eq.~\eqref{eqn-MexpSTchannel} and footnote \ref{ft6}, there may be additional entire functions or polynomial terms on the right-hand side. We neglect these contributions, as they do not affect our pole-based derivation.  }
\begin{equation}
A(S,T) = \sum_{\tau_s,\ell} C_s(\tau_s,\ell)\, A_{\tau,\ell}(S,T)\big|_{S\text{-pole}} +
\sum_{\tau_t,\ell} C_t(\tau_t,\ell)\, A_{\tau,\ell}(S,T)\big|_{T\text{-pole}}~,
\end{equation}
where \( A_{\tau,\ell}(S,T) \) denotes  the Borel-transformed Mack polynomials from each exchange operator. In particular, \( A_{\tau,\ell}(S,T) \) admits a strong-coupling expansion in powers of \( 1/\lambda^{1/4} \), which can be computed order by order. It contains poles at \( S,T = \delta \in \mathbb{Z}_{\ge 0} \).
On the other hand, the full amplitude \( A(S,T) \) also admits a worldsheet representation involving single-valued multiple polylogarithms, with a priori undetermined coefficients. By matching the residues at the poles of \( S \) and \( T \)  computed in these two methods, we are able to completely determine the coefficients corresponding to the leading curvature corrections.

Let us now analyze these contributions for both $S$-channel and $T$-channel poles  respectively.

\subsubsection{$S$-channel}

We now consider the $S$-pole:\footnote{To ease the notation, we use $\tau$ in replace of $\tau_s$ in this subsubsection.}
 \be
 A_{\tau, \ell}(S,T)|_{S-\mathrm{pole}}=\lambda^{\frac12}\Gamma(p+1)\int\frac{\dd\alpha}{2\pi \ii}\frac{e^\alpha}{\alpha^ { p+2}} 
 \sum_{m=0}^\infty \frac{\cQ_s\Big(\frac{2\sqrt\lambda U}{\alpha}  +\frac{2p}{3}; \tau , \ell, m\Big)}{\frac{2\sqrt\lambda S}{\alpha}+\frac{2p}{3}-\tau-2m}~,
 \ee
 where $\cQ_s $  is given in \eqref{cQs}.
 Picking the residue at
 \be
 \alpha_*=\frac{2\sqrt{\lambda}S}{\tau+2m -\frac{2p}{ 3}}~,
 \ee
 we get 
 \be
  A_{\tau, \ell}(S,T)|_{S\text{-pole}}= -\frac{\Gamma(p+1)}{2S} \sum_{m=0}^\infty \frac{e^{\alpha_*}}{\alpha_*^p}
\cQ_s\Big(\frac{2\sqrt\lambda U}{\alpha}  +\frac{2p}{3}; \tau , \ell, m\Big)~.
 \ee

We focus on stringy operators with large twist $\tau \sim \lambda^{1/4}$, for which the sum over $m$ can be approximated by an integral over the continuum variable $x = m/\tau^2$ \cite{Alday:2022uxp, Alday:2022xwz}. Introducing the rescaled twist $\widetilde{\tau} = \tau / \lambda^{1/4}$, in the strong coupling limit $\lambda \to \infty$, we find
\be \label{eqn-Schannel-A-pole}
\begin{aligned}
 A_{\tau, \ell}(S,T)|_{S\text{-pole}} & = -\Gamma(p+1)\frac{\tau^2}{2S}\int_0^\infty \dd x \; \frac{e^{\alpha_*}}{\alpha_*^p}
\cQ_s\Big(\frac{2\sqrt\lambda U}{\alpha}  +\frac{2p}{3}; \tau , \ell, x\tau^2\Big) \\
& = \frac {(-1)^{p+1} \Gamma(p+1) 4^{l+\tau } \sin ^2\left(\frac{\pi  \tau }{2}\right)}{  \lambda ^{\frac{p}{2}+\frac{3}{4}} (1+\delta_{\ell,0})}
   \sum_{i=0}^\infty \frac{R_{\tau,\ell}^{(s) i}(S,U)}{\lambda^{\frac{i}{4}}}~ .
\end{aligned}
\ee
The explicit expressions of $R_{\tau,\ell}^{(s) i}(S,U)$ are given as
  \be
  \begin{aligned} \label{eqn-RS}
      R_{\tau,\ell}^{(s) 0}(S,U) & =  \frac{16 S^{-p-1} T_l\left(\frac{2 U}{S}+1\right)}{\pi ^3 \left(\tilde{\tau }^3-4 S \tilde{\tau }\right)}\, , \\
      R_{\tau,\ell}^{(s) 1}(S,U) & = -\frac{8 S^{-p-1} \left((2 l+1) \tilde{\tau }^2+4 (2 l+3) S\right) T_l\left(\frac{2 U}{S}+1\right)}{\pi ^3 \tilde{\tau }^2 \left(\tilde{\tau }^2-4 S\right)^2}~,
  \end{aligned}
  \ee
where $T_\ell(x)$ denotes the Chebyshev polynomials of the $\ell$-th order. The expression for $R_{\tau,\ell}^{(s) 2}(S,U)$ is too long to display, but can be found in the attached $\mathtt{Mathematica}$ notebook.

 One can verify that 
\be
R_{\tau,\ell}^{(s) i}(S,U) =(-1)^\ell R_{\tau,\ell}^{(s) i}(S,T) ~, \qquad U=-S-T~.
 \ee
Since the $S$-channel involves two identical operators, the spin $\ell$ of the exchanged operator must be even. As a result, the relation simplifies to $R_{\tau,\ell}^{(s) i}(S,U) = R_{\tau,\ell}^{(s) i}(S,T)$.

The expression \eqref{eqn-Schannel-A-pole} contains an exponentially large factor $4^\tau$ for stringy operators with $\tau = \mathcal{O}(\lambda^{1/4})$ in the strong coupling limit. Since the Mellin amplitude remains finite at strong coupling, this large factor from $R_{\tau,\ell}^{(s), i}(S,U)$ must be compensated by the OPE coefficients. To make this behavior explicit, we redefine the OPE coefficients as
\be
  C_s({\tau,\ell})=\frac  {  (-1)^{p+1} \lambda ^{\frac{p}{2}+\frac{3}{4}} (1+\delta_{\ell,0})}{4^{l+\tau }\Gamma(p+1) \sin ^2\left(\frac{\pi  \tau }{2}\right)}  f_s({\tau,\ell})~,
  \ee
where $f_s(\tau,\ell) = \mathcal{O}(1)$. With this redefinition, the $S$-channel contribution is given by
  \be
  C_s({\tau,\ell}) A_{\tau, \ell}(S,T)|_{S-\text{pole}}=  f_s({\tau,\ell}) \sum_{i=0}^\infty \frac{R_{\tau,\ell}^{(s) i}(S,U)}{\lambda^{\frac{i}{4}}}~ .
  \ee

\subsubsection{$T$-channel}

Next, we switch to  the $T$-pole:\footnote{For convenience, we use $\tau$ to denote $\tau_t$ in this subsection.}
 \be
 A_{\tau, \ell}(S,T)|_{T-\mathrm{pole}}=\lambda^{\frac12}\Gamma(p+1)\int\frac{\dd\alpha}{2\pi \ii}\frac{e^\alpha}{\alpha^ { p+2}} 
 \sum_{m=0}^\infty \frac{\cQ_s\Big(\frac{2\sqrt\lambda U}{\alpha}  +\frac{2p}{3}; \tau , \ell, m\Big)}{\frac{2\sqrt\lambda T}{\alpha}+\frac{2p}{3}-\tau-2m}~,
 \ee
 where $\cQ_t $  is given in \eqref{cQs}.

The integration  and summation in the $T$-channel can be evaluated in the same manner as before. Picking up the $\alpha$-residues from the denominator and converting the sum over $m$ into an integral, we finally arrive at
\be
\begin{aligned}
 A_{\tau, \ell}(S,T)|_  {T-\text{pole}} 
& = \frac { \Gamma(p+1) 4^{l+\tau } \sin ^2\left(\frac{\pi  \tau }{2}+\frac{(p-1)\pi}{2}\right)}{  \lambda ^{\frac{p}{2}+\frac{3}{4}} (1+\delta_{\ell,0})}
   \sum_{i=0}^\infty \frac{R_{\tau,\ell}^{(t) i}(T,U)}{\lambda^{\frac{i}{4}}}~.
\end{aligned}
\ee
The explicit forms of the first few $R_{\tau,\ell}^{(t), i}(S,U)$ are:
  \be \label{eqn-RT}
  \begin{aligned}
      R_{\tau,\ell}^{(t) 0}(T,U) & = \frac{16 ~T^{-p-1} T_l\left(\frac{2 U}{T}+1\right)}{\pi ^3 \left(\tilde{\tau }^3-4 T \tilde{\tau }\right)}~ , \\  \\
      R_{\tau,\ell}^{(t) 1}(T,U) & = \frac{8 T^{-p-1} \left((2 \ell+3+2p^2-4p) \tilde{\tau }^2+4 (2 \ell+1-2p^2+4p) S\right) T_l\left(\frac{2 U}{T}+1\right)}{\pi ^3 \tilde{\tau }^2 \left(\tilde{\tau }^2-4 T\right)^2}~.
  \end{aligned}
  \ee
The expression for $R_{\tau,\ell}^{(t), 2}(S,U)$ is lengthy and is  given in the attached \texttt{Mathematica} notebook.

 As before, we also redefine  the OPE coefficients 
  \be
  C_t({\tau,\ell})=\frac  {   \lambda ^{\frac{p}{2}+\frac{3}{4}}(1+\delta_{\ell,0})}{ \Gamma(p+1) 4^{l+\tau } \sin ^2\left(\frac{\pi  \tau  }{2}+\frac{(p-1)\pi}{2} \right)}  f_t({\tau,\ell})~,
  \ee
  such  that $f_t = \cO(1)$. With this redefinition, the $T$-channel contribution is given by
  \be
  C_t({\tau,\ell}) A_{\tau, \ell}(S,T)|_{T\text{-pole}}=  f_t({\tau,\ell}) \sum_{i=0}^\infty \frac{R_{\tau,\ell}^{(t) i}(T,U)}{\lambda^{\frac{i}{4}}}~.
  \ee

\subsection{Structure  of curvature  expansion }

The AdS Virasoro-Shapiro amplitude admits an expansion in the ’t Hooft-like coupling $\lambda$, or equivalently  in the AdS curvature $R_{\text{AdS}} \sim \lambda^{\frac14} \ell_s$:  
\be \label{eqn-Aexp}
A(S,T)=\sum_{i=0}^\infty \frac{A^{(i)}(S,T)}{\lambda^{\frac{i}{2}}}~.
\ee
The leading term $A^{(0)}(S,T)$ corresponds to the flat space VS amplitude, while the next term $A^{(1)}(S,T)$ represents the first non-trivial curvature correction. Notably, there is no correction at order $\mathcal{O}(\lambda^{-\frac{1}{4}})$.

Since the functions $A^{(i)}(S,T)$ are observables in string theory, one would ideally compute them directly from worldsheet techniques. However, such computations remain out of reach due to technical challenges associated with formulating string theory in backgrounds with Ramond-Ramond fluxes. Instead, following the approach of \cite{Alday:2023jdk,Alday:2023mvu}, we will construct a worldsheet representation using an ansatz involving single-valued multiple polylogarithms (SVMPLs), and determine the coefficients by matching with CFT data.

In the previous section, by performing the Borel transformation of \eqref{eqn-MexpSTchannel} and focusing on the contributions from stringy operators, we obtained\footnote{Note that there may be degeneracy in the operator spectra, namely two or more operators sharing identical $\tau, \ell$. In that case, we are not able to compute their OPE coefficients independently, but only their average denoted as $\EV{f_{s,t}}$.  }
\be \label{eqn-A-CFTpart}
A(S,T)=\sum_{\tau_s,\ell}
 f_s({\tau_s,\ell}) \sum_{i=0}^\infty \frac{R_{\tau_s,\ell}^{(s) i}(S,U)}{\lambda^{\frac{i}{4}}}
+
 \sum_{\tau_t,\ell}
 f_t({\tau_t,\ell}) \sum_{i=0}^\infty \frac{R_{\tau_t,\ell}^{(t) i}(T,U)}{\lambda^{\frac{i}{4}}}~.
\ee
To obtain the individual terms in the expansion \eqref{eqn-Aexp}, one has to expand the right-hand side of \eqref{eqn-A-CFTpart} in powers of $1/\lambda^{\frac12
}$. Since both the twist $\tau$ and the OPE coefficients $f_{s,t}(\tau,\ell)$ depend non-trivially on $\lambda$, we expand them as
 \beqn\label{taufexpansion}
 \tau_s &=&\tau^{(s)}_0 \lambda^{\frac{1}{4}}+\tau^{(s)}_1 
 +\tau^{(s)}_2\lambda^{-\frac{1}{4}}+\cdots~,
\\ \label{eqn-TauExp}
 \tau_t&=&\tau^{(t)}_0 \lambda^{\frac{1}{4}}+\tau^{(t)}_1 
 +\tau^{(t)}_2\lambda^{-\frac{1}{4}}+\cdots~,
\\
f_s(\tau,\ell)&=&f^{(s)}_0 +f^{(s)}_1 \lambda^{-\frac{1}{4}}
 +f^{(s)}_2 \lambda^{-\frac{1}{2}}+\cdots~,
 \\
 f_t(\tau,\ell)&=&f^{(t)}_0 +f^{(t)}_1 \lambda^{-\frac{1}{4}}
 +f^{(t)}_2 \lambda^{-\frac{1}{2}}+\cdots~,
\eeqn
where all coefficients $\tau^{(s,t)}_i$ and $f^{(s,t)}_i$ are $\lambda$-independent.

\subsection{The flat limit: $\cO(\lambda^0)$}

The leading-order term  $\cO(\lambda^0)$ in the   expansion of \eqref{eqn-A-CFTpart} is given explicitly by
\be \label{flatVS-cft}
A^{(0)}(S,T) = f^{(s)}_0  R_{\tau_0^{(s)}\lambda^{\frac{1}{4}},\ell}^{(s) 0}(S,U) +f_0^t R_{\tau_0^{(t)}\lambda^{\frac{1}{4}},\ell}^{(t) 0}(T,U) ~.  
\ee
This is expected to coincide with the flat space Virasoro-Shapiro amplitude  \eqref{flatVS} obtained from the worldsheet integral\footnote{Here the measure is defined as $\dd^2z= \dd z \dd \bar z /(-2\pi \ii)$.}
\begin{equation}
A^{(0)}(S, T) = \int \,{\dd^2 z  \; |z|^{-2S-2}}{|1 - z|^{-2T-2}}
\frac{S^2+T^2+U^2}{12U^2} 
+ (S \leftrightarrow T)+ (S \leftrightarrow U )~,
\end{equation} 
where $S+T+U=0$.
Explicitly, it can be evaluated  as 
\be 
A^{(0)}=-\left(S^2+T^2+U^2\right)\frac{\Gamma (-S) \Gamma (-T) \Gamma (-U)  }{4 \Gamma (S+1) \Gamma (T+1) \Gamma (U+1)}~, \qquad  S+T+U=0~ .
\ee

The flat space VS amplitude $ A^{(0)}(S,T)$ \eqref{flatVS} has simple poles at $S = \delta \in \mathbb{Z}_{\geq 0}$ and $T=\delta \in \mathbb{Z}_{ \geq 0}$. By matching the residues of   $A^{(0)}$ at these poles with those obtained from the conformal correlator expression \eqref{flatVS-cft}, we are able to determine the leading order OPE data as follows:
\begin{itemize}
    \item $\tau_0^{(s)}(\delta, \ell) = \tau_0^{(t)}(\delta, \ell) = 2\sqrt{\delta}$.
    \item $\langle f_0^{(s)}(\delta, \ell) \rangle$ and $\langle f^{(t)}_0(\delta, \ell) \rangle $ are non-vanishing if and only if $\ell \in [0,2\delta]$. 
    Some values of those OPE coefficients are shown in the attached $\mathtt{Mathematica}$ notebook.
    \item The leading Regge trajectory has $\delta = \ell/2$ with even $\ell$, starting from $\ell =2$.
\end{itemize}
 For operators on the leading trajectory, we find
\be \label{eqn-f0Leading}
 f_0^{(s)}(\delta, 2\delta)= f_0^{(t)}(\delta, 2\delta) = f_0 (\delta, 2\delta) \equiv \frac{\pi^3\delta ^{2 \delta +p-\frac{1}{2}}}{2^{4 \delta +1} \Gamma (\delta )^2}\, .
\ee

One can also understand the large anomalous dimension from holography intuitively as follows.   
Holographically the  long multiplets of relevance here are related to massive string states. In the infinite coupling limit, their twist \( \tau  \) is given by \( mR_{\text{AdS}} \), where \( m \) is the mass of the string state and \( R_{\text{AdS}} \) is the radius of AdS. In type IIB string theory, the string energy levels are \( m^2 = \frac{4\delta}{\alpha'} \), with \( \delta \in \mathbb{N} \). As a consequence
\begin{equation}
\tau  \xrightarrow{\lambda\to\infty}  2\sqrt{\delta}\,\lambda^{\frac{1}{4}}~, \quad \delta \in \mathbb{N}~, 
\end{equation}
where recall that the 't Hooft-like coupling is related to the AdS radius   as \( \lambda =\frac{R_{\text{AdS}}^4}{\alpha'^2} \).
 Comparing to \eqref{taufexpansion}, we immediately learn that $\tau_0=2\sqrt{\delta}$ for both $S$ and $T$ channels. 

\subsection{Subleading order $\mathcal{O}(\lambda^{-\frac{1}{4}})$}

At subleading order $\mathcal{O}(\lambda^{-\frac{1}{4}})$, we expect the AdS VS amplitude to vanish, since physical processes are organized as a series expansion in powers of $1/\sqrt{\lambda}$. However, the conformal data, particularly the twist expansion of stringy operators given in \eqref{eqn-TauExp}, starts at order $\mathcal{O}(\lambda^{\frac{1}{4}})$. As a result, the CFT contribution to \eqref{eqn-A-CFTpart} at this order is not manifestly zero.

To ensure consistency with the expected vanishing of the AdS amplitude, we use the explicit expressions for the residues in \eqref{eqn-RS} and \eqref{eqn-RT}, and impose the cancellation of the $\mathcal{O}(\lambda^{-\frac{1}{4}})$ term. This leads to the following constraints on the subleading twist corrections and OPE coefficients:
\be
\begin{aligned}
\tau_1^{(s)} & = \tau_1^{(t)} = -\ell -1~, \\    
 \langle f_1^{(s)}(\delta ,\ell) \rangle & = -\frac{(4 \ell +5) }{4 \sqrt{\delta }} \langle f_0^{(s)}(\delta ,\ell) \rangle~, \quad \langle f_1^{(t)}(\delta ,\ell) \rangle = -\frac{\left(4 \ell-2 p^2+4 p+3\right)}{4 \sqrt{\delta }} \langle f_0^{(t)}(\delta ,\ell) \rangle~.
\end{aligned}
\ee
For $p=1$, these results reproduce those of \cite{Chester:2024wnb}. For $p > 1$, the $p$-dependence of the subleading OPE coefficients $f_1^{(s)}$ and $f_1^{(t)}$ differs: notably, the $S$-channel coefficient $f_1^{(s)}$ is independent of $p$. This is expected, as the $S$-channel involves the OPE of $  \cO_1 \cO_1  $, which is identical to the $p=1$ case regardless of the remaining two operators.

\subsection{The first curvature correction and subsubleading order $\cO(\lambda^{-\frac12})$}

The first non-trivial correction appears at order $\mathcal{O}(\lambda^{-\frac{1}{2}})$. In contrast to the leading-order result \eqref{flatVS}, the corresponding worldsheet expression for this correction is not known explicitly. However, as demonstrated in \cite{Alday:2023jdk,Alday:2023mvu,Fardelli:2023fyq}, a physically motivated worldsheet ansatz for $A^{(1)}$ can be made, and it proves sufficient to determine the final result.

\subsubsection{Worldsheet ansatz}

For the correlator $\langle pp11 \rangle$, the amplitude at order $\mathcal{O}(\lambda^{-\frac{1}{2}})$ satisfies the crossing relation $A^{(1)}(S,T) = A^{(1)}(S,U)$. This symmetry is made manifest by the following ansatz:
\be\label{A1BD}
A^{(1)}(S,T)=B(S,T)+B(S,U)+B(U,T)+D(S,T)+D(S,U)~,
\ee
where
\be
B(S,T)=\int \dd^2 z\; |z|^{-2S-2} |1-z|^{-2T-2}G(S,T,z)~, 
\ee
and
\be
D(S,T)=\int \dd^2 z\; |z|^{-2S-2} |1-z|^{-2T-2}H(S,T,z)~.
\ee
The insertions $G$ and $H$  on the worldsheet are linear combinations of a special set of functions, to be specified in detail below.

The idea behind this ansatz is as follows: the amplitude $A^{(1)}$ contains a fully crossing-symmetric part under $S \leftrightarrow T \leftrightarrow U$, which is captured by the $B$-terms. In contrast, the $D$-terms are symmetric only under $T \leftrightarrow U$. When $p=1$, all four operators are identical and the correlator is fully crossing symmetric under permutations of $S,T,U$. Therefore, we expect that
\be
D(S,T)=0~, \qquad\text{if}\qquad p=1~  .
\ee

By changing integration variables, one can show that
\beqn
B(U,T)&=&\int \dd^2 z\; |z|^{-2S-2} |1-z|^{-2T-2} \times |z|^2\, G\left(U,T,1/z \right)~, \\
B(S,U)&=&\int \dd^2 z\; |z|^{-2S-2} |1-z|^{-2T-2} \times |1-z|^2\, G\left(S,U,\frac{z}{z-1} \right)~,\qquad
\eeqn
and similarly for the $D$ terms. As a result, the full amplitude can be written as a single worldsheet integral:
\be \label{eqn-A1-worldsheet}
A^{(1)}(S,T)=\int \dd^2 z\; |z|^{-2S-2} |1-z|^{-2T-2}G_{\mathrm{tot}}(S,T,z)~, 
\ee
where 
\beqn
G_{\mathrm{tot}}(S,T,z)&=&G (S,T,z)+|z|^2 G(U,T,1/z)+|1-z|^2 G(S,U,\frac{z}{z-1}) \nonumber
\\&&
  +H (S,T,z)+|1-z|^2 H(S,U,\frac{z}{z-1})~.
\eeqn

Following \cite{Alday:2023jdk,Alday:2023mvu,Fardelli:2023fyq}, we assume that the integrands $G$ and $H$ can be expressed as linear combinations of a set of special functions known as single-valued multiple polylogarithms (SVMPLs), denoted by $\mathcal{L}_w(z)$. Here, the label $w$ refers to a word formed from the alphabet ${0,1}$, and the weight of the SVMPL corresponds to the length of the word. For the problem at hand, we need to use the weight three SVMPL to write down the worldsheet ansatz.    The explicit form of the weight three SVMPL can be found in appendix \ref{apd-SVMPL}.
To make the crossing properties   manifest, we will use the following equivalent basis 
\be \label{SVMPLbasis}
 \mathcal{L}^s =\Big(\mathcal{L}_{000}^s, \mathcal{L}_{001}^s, \mathcal{L}_{010}^s, \zeta(3)\Big)~,  \qquad \mathcal{L}^a =\Big(\mathcal{L}_{000}^a, \mathcal{L}_{001}^a, \mathcal{L}_{010}^a\Big)~ ,
\ee
where
  \be
 \begin{aligned} & \mathcal{L}_w^s(z)=\mathcal{L}_w(z)+\mathcal{L}_w(1-z)+\mathcal{L}_w(\bar{z})+\mathcal{L}_w(1-\bar{z})~, \\ & \mathcal{L}_w^a(z)=\mathcal{L}_w(z)-\mathcal{L}_w(1-z)+\mathcal{L}_w(\bar{z})-\mathcal{L}_w(1-\bar{z}) ~.
 \end{aligned}
 \ee
The functions $\mathcal{L}_w^s$ ($\mathcal{L}_w^a$) are symmetric (resp. antisymmetric) with respect to $z \leftrightarrow 1-z$.\footnote{Note $\cL_w(\bar z) =\cL_{\tilde w}(z)$ where $\tilde w$ denotes the reversed word of $w$. For instance, $\widetilde{001}=100$. }

 Using the basis \eqref{SVMPLbasis}, we can  write $G$ and $H$ as
\be \label{eqn-A1-GHansatz}
\begin{aligned}
G(S,T,z)&=\sum_{j=1}^4 g^s_j (S,T)\cL^s_j(z)+\sum_{j=1}^3 g^a_j (S,T)\cL^a_j(z)~,
\\
H(S,T,z)&=\sum_{j=1}^4 h^s_j(S,T) \cL^s_j(z)+\sum_{j=1}^3 h^a_j(S,T) \cL^a_j(z)~,
\end{aligned}
\ee
where the coefficients $g^s_j$, $g^a_j$, $h^s_j$, and $h^a_j$ are homogeneous polynomials of degree two in $S$ and $T$, of the general form $a S^2 + b T^2 + c ST$. The coefficients $a$, $b$, and $c$ are themselves functions of the external   $p$.

Additional constraints arise from symmetry considerations. In particular, the crossing symmetry
\be
B(S,T)=B(T,S) \longleftrightarrow G(S,T,z)=G(T,S,1-z)~
\ee
implies that the symmetric coefficients $g^s_j(S,T)$ must be symmetric in $S$ and $T$ (i.e., $a = b$), while the antisymmetric coefficients $g^a_j(S,T)$ must be antisymmetric (i.e., $a = -b$ and $c = 0$). In total, there are $4\times (2+3)+3\times (1+3)=32$ coefficients to be fixed.

\subsubsection{Integration over the worldsheet}

To get the final result and compare with CFT, the next step is performing the worldsheet integral involving SVMPL. A typical integration takes the following form
\be
  I_w(S,T)=\int \dd^2 z\; |z|^{-2S-2}|1-z|^{-2T-2}\cL_w(z)~.
\ee
By Taylor expanding in $S,T$, one can show that it takes the  form \cite{Alday:2023jdk}\footnote{Here $\shuffle$ denotes the   shuffle product,  which  combines two words by merging them together while preserving the original orders.}
 \be
  I_w(S,T)=\mathtt{polar}_w+\sum_{p,q=0}^\infty (-S)^p (-T)^q \sum_{W \in 0^p \shuffle 1^q \shuffle w}(\cL_{0W}(1) -\cL_{1W}(1))~,
  \ee
  where $\mathtt{polar}_w$ means singular terms in $S,T$, namely poles of the form $S^p T^q$ for $p<0$ or $q<0$.  

The polar terms arise from specific regions of the complex $z$-plane. In particular, all singular terms of the form $\#/(S - \delta)^i$ with $i = 1,2,3,4$ originate from the region where $|z| \to 0$. To isolate these contributions, we expand the integrand around small $|z|$ using polar coordinates $z = \rho e^{i\theta}$ and perform a Taylor expansion in $\rho$ near $\rho = 0$. Integrating term by term over $\theta$ and then over $\rho$, we obtain contributions of the form $\rho^{-1 - 2S + 2\delta} \log^p \rho^2$. The integral evaluates as \cite{Alday:2023jdk,Alday:2023mvu}
  \be
  \int_0^ {\rho_0}  \dd \rho\;   \rho^{-2S+2\delta -1} \log^p \rho^2  =-\frac{p!}{2}\frac{1}{(S-\delta)^{p+1}}+ \cO((S-\delta)^0)~,
   \ee
where the dependence on the cutoff $\rho_0$ only enters the regular terms. Similarly, the $T$-channel poles can be extracted by expanding around the region $z \to 1$.

\subsubsection{Results}

By matching the coefficients of all poles in the worldsheet integral \eqref{eqn-A1-worldsheet} with those in the CFT expression \eqref{eqn-A-CFTpart}—specifically, the $S$-channel poles at $S = \delta = 1,2,\ldots$ and the $T$-channel poles at $T = \delta = 1,2,\ldots$—and further supplementing with the supergravity (SUGRA) results at $S = T = 0$, we fully determine all coefficients in the worldsheet ansatz, and thereby fix the CFT data at this order.

The final expression of the worldsheet integrand is fully determined  up to five ambiguities (detailed in appendix \ref{zeromd}) whose integrals vanish. With an appropriate choice of these ambiguities, the worldsheet coefficients in \eqref{eqn-A1-GHansatz} take the form: 
\begin{subequations}\label{ghcoef}
\begin{align} 
g^{s}_j= &\Big(\frac{1}{12} \left(2 p-  p^2\right) S T,\ \frac{1}{48} \left(2 p^2-2 p+9\right) \left(S^2+T^2\right)+\frac{1}{24} \left(p^2-3 p\right) S T, \nonumber \\
& \;\; \frac{5}{96} \left(-  p^2+  p-6\right) \left(S^2+T^2\right)+\frac{1}{48} \left(p^2-3 p-6\right) S T,\  \nonumber  \\
&\;\; \frac{1}{6} \left(2 p^2-2 p+15\right) \left(S^2+T^2\right)\Big)~ ,
          \\[1ex] 
           g^a_j = & \Big(0,\ \ \frac{1}{48} \left(-2 p^2+6 p-3\right) \left(S^2-T^2\right),\ \ \frac{1}{96} \left(-p^2+5 p+6\right) \left(S^2-T^2\right) \Big)~,
         \\[1ex] 
         h^s_j  = &  \frac{(  p-1)p}{96} 
   \Big( -S^2+6 S T-2 T^2,\ \ 0,\ \ 2 S^2+T^2,\ \ 0 \Big)~,
           \\[1ex]
 h_j^a = &   \frac{(  p-1)p}{96} 
  \Big( -S^2+6 S T+2 T^2,\ \ -8 S T,\ \ 2 S^2-4 S T-T^2\Big)~ .
\end{align}
\end{subequations}
As expected, the coefficients  $ h_j^{s/a}$ of the $D$ term vanish  when $p=1$.

Inserting the solution to  \eqref{A1BD} and doing the worldsheet integration, we get the low energy expansion of the AdS Virasoro-Shapiro amplitude: 
\beqn
A^{(1)}&=& -\frac{p}{12}   \left( \frac{(p-3)}{T^2}+\frac{(p-3)}{U^2} -\frac{2 p}{S^2}\right)+\frac{1}{2} (p-1) p S \zeta (3) \nonumber\\
&&
+\zeta (5) \left((p-1) p S^3-\frac{1}{4} (4 p (p+2)+45) S T U\right)
\nonumber \\  
&&
+\frac{1}{6} \zeta (3)^2 \left((p^2-9p -54) S^4+4 (p^2+3p +27) S^2 T U-2 (p^3 + 3p +27) T^2 U^2\right)+\cdots  ~.\qquad\qquad 
\eeqn

\paragraph{Anomalous dimensions $\tau_2$:}

We begin by presenting the CFT data for the scaling dimensions of the exchanged operators. For operators lying on the leading Regge trajectory, we find:
\be \label{eqn-pp11-tau2}
\tau_2^{(s)}\left(\frac{\ell}{2},\ell \right) = \frac{3\ell^2 - 2\ell +2}{4\sqrt{2}\sqrt{\ell}}~, \quad \tau_2^{(t)} \left(\frac{\ell}{2},\ell \right) = \frac{3\ell^2 - 2\ell +2 p^2}{4\sqrt{2}\sqrt{\ell}}~,
\ee
where we recall $\ell \in 2\mathbb{Z}_{\geq 0}$.
As expected, the S-channel result is independent of $p$, reflecting the fact that this channel corresponds to the OPE limit $\cO_p \cO_p \to \cO_1 \cO_1$, where the exchanged operator structure matches that of the $p = 1$ case. Note that the $T$-channel exchanged long multiplet transforms under spin-($\frac{ p-1}{2}$,$\frac{ p-1}{2}$) representation under R-symmetry.

So far, there are no alternative independent checks—particularly from quantum integrability methods—for the results presented above. The only available cross-check comes from comparison with the semiclassical folded string solution, which carries quantum numbers $(\mathcal{S}, J_1, J_2) = (\ell, J, 0)$, as detailed in \cite{Chester:2024wnb}. The basic idea of this method is to start in a regime where both the spin $\ell$ and the R-charge $J$ of the corresponding string state are large, $\sim \mathcal{O}(\sqrt{\lambda})$, so that quantum fluctuations are suppressed. After performing the semiclassical computation in this limit, one then extrapolates the result to the regime of interest, where $\ell, J = \mathcal{O}(1)$.\footnote{The extrapolation can be subtle. For instance, in the $\AdS_4\times \mathbb{CP}^3$ case, the $\frac{J^2}{2\sqrt{2\ell}}$ term receives quantum corrections and becomes $\frac{(J+1/2)^2}{2\sqrt{2\ell}}$, as discussed in \cite{Gromov:2014eha}. In the $\AdS_5\times S^5$ case and also in our case, there are no such additional shifts.}

The semiclassical result for the energy of the folded string is given by
\be \label{eqn-semiAns}
E =\lambda^{\frac14}\sqrt{2\ell}+\lambda^{-\frac14}\times \frac{3\ell^{2} +8 \cC \ell +2J^2}{4\sqrt{2} \ell} +\cdots
~,
\ee
where $\mathcal{C}$ denotes a one-loop coefficient that has not yet been computed.\footnote{For the case $M_4 = T^4$, one-loop determinant computations around the semiclassical folded string solution have been performed in \cite{Beccaria:2012kb,Beccaria:2012pm}. However, due to unresolved regularization issues, a definitive one-loop result could not be established.} This expression matches our result for $\tau^{(s)}_2$ (respectively $\tau^{(t)}_2$) in \eqref{eqn-pp11-tau2}, upon setting $J = 1$ (respectively $J = p$) and identifying the semiclassical string states as descendants obtained by acting with two supercharges on the exchanged primary operators, as discussed in \cite{Chester:2024wnb}.

Our result also yields a prediction for the one-loop coefficient:
\be
\cC=-{1}/{4}~,
\ee
which coincides with the value previously obtained for the $\AdS_5 \times S^5$ background in \cite{Roiban:2011fe}.

Operators on higher Regge trajectories are degenerate; therefore, the results obtained from our method correspond to averaged expressions over the degenerate states. The explicit results are provided in the attached \texttt{Mathematica} notebook.

\paragraph{OPE coefficients $f_2$:}

The OPE coefficients in both $S$ and $T$ channels exhibit explicit $p$-dependence.

For operators on the leading Regge trajectory (i.e., with spin $\ell = 2\delta$),
we find  the following analytic expressions for the  
OPE coefficients:
\be
\begin{aligned}
\frac{f^{(s)}_2\left(\delta,2\delta\right)}{f _0\left(\delta,2\delta\right)} & = \frac{21}{32 \delta}-\frac{7 \delta^2 }{6}+\frac{2}{3 }+\frac{15\delta}{4}+\frac{\left(3 \delta ^2-3 \delta +2\right) p}{6 \delta} -p^2-\frac{p^3}{3 \delta }+ 2\delta^2\, \zeta(3)\, ,\\  \frac{f^{(t)}_2\left(\delta,2\delta\right)}{f _0\left(\delta,2\delta\right)}  & = -\frac{7 \delta ^2}{6}+\frac{15 \delta }{4}+\frac{1}{32 \delta }-\frac{5}{6} +p\left(\frac{\delta }{2}+\frac{7}{12 \delta }+\frac{3}{2}\right)  +p^2 \left(\frac{1}{4 \delta }-\frac{3}{2}\right) \\ 
& \hspace{4ex} -\frac{p^3}{3 \delta } +\frac{p^4}{8 \delta }+ 2\delta^2\, \zeta(3)\, ,
\end{aligned}
\ee
where the leading-order OPE coefficients $f_0(\delta, 2\delta)$ are given analytically in \eqref{eqn-f0Leading}. More explicitly,  the first few OPE coefficients on the leading Regge trajectory are given by:
\be
\begin{aligned}
    f^{(s)}_2\left(1,2\right) & = \pi^3 \frac{-32 p^3-96 p^2+32 p+375}{3072} + \frac{\pi^3
   }{16} \zeta(3)~, \\
f^{(s)}_2\left(2,4\right)& = \pi^3 \frac{2^{p-\frac{23}{2}} }{3} \left(-32 p^3-192 p^2+128 p+735\right) + \pi^3 2^{p-\frac{5}{2}} \zeta(3)~, \\
    f^{(t)}_2\left(1,2\right) & = \pi^3 \frac{12 p^4-32 p^3-120 p^2+248 p+171}{3072} + \frac{\pi^3}{16} \zeta(3)~, \\
    f^{(t)}_2\left(2,4\right) & = \pi^3 \frac{2^{p-\frac{23}{2}}}{3}  \left(12 p^4-32 p^3-264 p^2+536 p+387\right) + \pi^3 2^{p-\frac{5}{2}} \zeta(3)~ . \\
\end{aligned}
\ee

More OPE data can be found in the accompanying \texttt{Mathematica} notebook. As previously explained, operators on higher Regge trajectories are degenerate, and our results represent averaged expressions over these degenerate states. 

\section{Virasoro-Shapiro amplitude for  $\EV{p_1p_2p_3p_4}$}
\label{wsp1234}

 In this section, we will generalize the study of $\EV{pp11}$ correlators to the most general case   $\EV{p_1p_2p_3p_4}$
 with four  arbitrary KK modes, utilizing the $\AdS\times S$ formalism.

\subsection{$\AdS\times S$ formalism for Mellin amplitude and its Borel transformation }

  In the previous sections, we consider the Mellin amplitudes which are obtained from the Mellin transformation of the correlation functions in the position space of boundary CFT. 
To account for KK modes, it is convenient to perform an additional Mellin transform on the internal $R$-symmetry space $S^3$, as introduced in \cite{Aprile:2020luw,  Aprile:2021mvq}. The resulting object is known as the $\AdS \times S$ Mellin amplitude, which in  the current case is defined as
\be\label{AdSSMellin}
M(\sfy_{ij})=\sum_{m_{ij}\in \bZ}\cM(m_{ij} ) \prod_{1\le i<j\le 4}\frac{\sfy_{ij}^{m_{ij}}}{\Gamma(m_{ij}+1)}~,
\ee
where
\be\label{mijdiff}
\sum_{i}m_{ij}=0~, \qquad m_{ii}= 1-p_i~, \qquad m_{ij}=m_{ji}~.
\ee
Note that the infinite sum above actually truncates to a finite sum due to the Gamma function in the denominator.
The constrained system \eqref{mijdiff} over $m_{ij}$ contains two free parameters, which can be chosen as $m_{24},m_{34}$. The final expression of the $\AdS \times S$ Mellin amplitude   \eqref{AdSSMellin} can be shown to be  
\be\label{Msmij}
   M(s,t,\sigma, \tau )=\sum_{m_{24},m_{34}\ge 0\atop  m_{24}+m_{34}\le L-1}
\frac{   \sigma^{m_{24}}\tau^{L-1-m_{24}-m_{34}}  
}{  \prod_{i<j}\Gamma(m_{ ij}+1) }
\cM(s,t;m_{ij})~,
\ee
 where the LHS is given in   \eqref{Mellin}.
 
In the supergravity limit, the $\AdS \times S$ Mellin amplitude takes a remarkably simple form\footnote{The explicit expression  of $\cM_{\text{SG}}$ in \eqref{Msmij} was computed in \cite{Wen:2021lio} and also discussed in \cite{Behan:2024srh}, although they were not using the $\AdS\times S$ formalism explcitily.   }\footnote{The property that the Mellin amplitude in the supergravity limit depends only on specific combinations such as \( s + m_s - \Sigma \) is related to the hidden conformal symmetry of correlators on \(\AdS_3 \times S^3\).
}
\be \label{sugraM2}
  \cM_{\text{SG}}(s,t;m_ {ij})=-\frac{1}{\bm s+2}-\frac{1}{\bm t+2}-\frac{1}{\bm u+2}~,
  \ee
 where the boldfaced Mandelstam variables are defined as:
   \beqn \label{boldstu}
   \bm s &=& s-p_1-p_2+2 m_{12}=s-p_3-p_4+2 m_{34}=s-   \Sigma  +m_s~ ,
\\
   \bm t &=& t-p_1-p_4+2 m_{14}=t-p_2-p_3+2 m_{23}
   =t-   \Sigma  +m_t~,
\\
   \bm u &=& \tilde  u-p_1-p_3+2 m_{13}=\tilde  u-p_2-p_4+2 m_{24}
   =\tilde  u-    \Sigma  +m_u\, .
   \eeqn
Here, we introduced:
   \be\label{mstu}
   m_s=m_{12}+m_{34}~,  \quad
   m_t=m_{14}+m_{23}~,  \quad
   m_u=m_{13}+m_{24}~, \quad
   \Sigma=\frac12(p_1+p_2+p_3+p_4)~ .
   \ee   
   
These quantities satisfy the constraints:   
\be
m_s+m_t+m_u
=\Sigma-2~, \qquad
s+t+\tilde u =2\Sigma-2~,
\ee
which imply:
   \be
   \bm s+\bm t+\bm u=-4~ .
   \ee

As in  the  case of Mellin amplitude, we can furthermore perform a Borel transformation on the $\AdS\times S$ Mellin amplitude \eqref{Aborel} and define 
\be
\cA(S,T;m_ {ij})=\lambda^{\frac12}\Gamma(  \Sigma) \int\frac{\dd\alpha}{2\pi \ii}\frac{e^\alpha}{\alpha^ {\Sigma+1}} 
\cM\Big(\frac{2\sqrt\lambda S}{\alpha}+\frac{2\Sigma -2}{3},\frac{2\sqrt\lambda T}{\alpha}+\frac{2\Sigma -2}{3};m_ {ij}\Big)~.
 \ee

 When applied to the supergravity part \eqref{sugraM2}, the Borel transform gives the following  supergravity contribution  
\be
\cA_{\text{SG}} = \frac{S^2+T^2+U^2}{ 4 S T U }
-\frac{  {  \Sigma } -1 }{12\sqrt{\lambda }}
\Big[     \frac{\Sigma-3   {m_s}-4}{S^2}+
\frac{\Sigma-3   {m_t} -4}{T^2}
+
\frac{\Sigma-3 m_u -4}{U^2}
\Big]+ \cO(1/\lambda)~.
\ee

\subsection{Worldsheet ansatz}

Our next goal is to find a worldsheet representation of the AdS Virasoro--Shapiro amplitude \(\cA\) for four arbitrary KK modes \(p_i\). As discussed above, in the \(\AdS \times S\) framework, both \(\cM\) and \(\cA\) are functions of the \(m_{ij}\), or equivalently, of the two independent variables \(m_{24}, m_{34}\). For simplicity and symmetry, we will further assume that \(\cM\) and \(\cA\) depend on \(m_{24}, m_{34}\) only through the combinations \(m_s, m_t\) defined in \eqref{mstu}. Thus, we assume  the functional forms \(\cM(S, T; m_s, m_t)\) and \(\cA^{(1)}(S, T; m_s, m_t)\), where we suppress the dependence on \(\Sigma\). This assumption ensures manifest symmetry between the \(\AdS\) and \(S\) spaces and is consistent with the supergravity result \eqref{sugraM2}.

With this assumption, we can now  make the following worldsheet ansatz
\be\label{cA1p1234}
\cA^{(1)}(S,T;m_s,m_t)=\cB(S,T; m_s,m_t)+\cB(S,U; m_s, m_u)+\cB(U,T; m_u, m_t) ~,
\ee
where
\be
\cB(S,T; m_s,m_t)=\int \dd^2 z\; |z|^{-2S-2} |1-z|^{-2T-2}\cG(S,T,z; m_s,m_t)~, 
\ee
and
\be\label{cGcoeff}
\cG(S,T,z; m_s,m_t)=\sum_{j=1}^4  \cR_j^s(S,T;m_s,m_t)\cL_j^s(z)+\sum_{j=1}^3  \cR_j^a(S,T;m_s,m_t)\cL_j^a(z)~.
\ee 

As in the previously studied case of $\EV{pp11}$, we make the further assumption that the functions $\cR_j^{s/a}(S,T; m_s, m_t)$ are given by homogeneous polynomials of degree two in \(S\) and \(T\), with coefficients given by the linear combinations of\footnote{Since there is a constraint $m_s+m_t+m_u=\Sigma-2$, the number of bases is actually reduced and can be chosen as
 $ 
 \{\Sigma m_s, \Sigma m_t, \Sigma, \Sigma^2, m_s,m_t \} ~.
 $}  
\be
\{\Sigma m_s, \Sigma m_t, \Sigma m_u, \Sigma, m_s, m_t, m_u\} ~. 
\ee
 
In total, there are  $(4+3)\times 7\times 3=147$ coefficients to be fixed.
 
 \subsection{Fixing coefficients}
  We now impose the crossing symmetry : 
\be
\cB(S,T ; m_s,m_t)=\cB(T,S; m_t,m_s)~,
\ee
leaving 75 coefficients unfixed. 

To further fix these coefficients, we can compare with the  case of  $\EV{pp11}$  \eqref{A1BD} studied in the last section. In this case,  $m_s=p-1, m_t=m_u=0$. As a result, we should have:\footnote{Note that $H(U,T)$ is absent in the last equation above. Equivalently, one can think that their coefficients are zero in this case. }
\beqn
\cG(S,T,z; m_s=p-1,m_t=0)&=&G(S,T,z)+H(S,T,z)~, \\
\cG(S,U,z; m_s=p-1,m_u=0)&=&G(S,U,z)+H(S,U,z)~, \\
\cG(U,T,z; m_u=0,m_t=0)&=&G(U,T,z ) ~, 
\eeqn
where the equality above means that coefficients in the basis of $\cL_j^{s/a}(z)$  are the same.   

 Using the explicit result in \eqref{ghcoef},  a direct comparison of coefficients gives the final result
 
\beqn\renewcommand{\arraystretch}{1.5}
\label{RRscof}
\cR^s=\footnotesize
\begin{pmatrix}
\frac{1}{192} \left(2 S T (2 + 9 (m_s  + m_t )(-1 + \Sigma) + 20 \Sigma - 4 \Sigma^2) + 3 (S^2 +T^2 )(5   - 3 \Sigma + \Sigma^2)  - 6( m_s T^2+m_t S^2)(\Sigma-1)\right) \\
\frac{1}{48} \left(-3 m_t S^2 (-1 + \Sigma) - T (3 m_s T (-1 + \Sigma) + S (7 + \Sigma + \Sigma^2))\right) \\
\frac{1}{92} \left(6 T^2 (-2 + m_s (-1 + \Sigma)) + 6 S^2 (-2 + m_t (-1 + \Sigma)) - S T (19 + \Sigma + \Sigma^2)\right) \\
\frac12 T^2 (2 + m_s - m_s \Sigma) + \frac12 S^2 (2 + m_t - m_t \Sigma)  
\end{pmatrix}~,\qquad\quad 
\eeqn

  \beqn\renewcommand{\arraystretch}{1.5}
  \label{RRacof}
  \cR^a=\footnotesize
\begin{pmatrix}
\frac{1}{64} \left(6 (m_s - m_t) S T (-1 + \Sigma) + T^2 (-5 + 2 m_s (-1 + \Sigma) + 3 \Sigma - \Sigma^2) + S^2 (5 - 2 m_t (-1 + \Sigma) - 3 \Sigma + \Sigma^2)\right) \\
\frac{1}{48} \left(-6 (m_s - m_t) S T (-1 + \Sigma) - T^2 (2 + 3 m_s (-1 + \Sigma) + 4 \Sigma) + S^2 (2 + 3 m_t (-1 + \Sigma) + 4 \Sigma)\right) \\
\frac{1}{48} \left(-3 (m_s - m_t) S T (-1 + \Sigma) + S^2 (1 + 2 \Sigma) - T^2 (1 + 2 \Sigma)\right)
\end{pmatrix}~.\qquad\quad 
\eeqn

With this solution, we can compute the low energy expansion of the $\AdS\times S$ Virasoro-Shapiro amplitude:
  \beqn
  \cA^{(1)}(S,T;m_s,m_t)&=&
  \frac{1-\Sigma}{12}   \left(\frac{-3 m_s+\Sigma -4}{S^2}+\frac{-3 m_t+\Sigma -4}{T^2}+\frac{-3 m_u+\Sigma -4}{U^2}\right)
   \nonumber \\&&
  +\frac{1}{2} (\Sigma -1)  (m_s S+m_t T+m_u U)\zeta (3) 
  \nonumber
  \\&&
 + \Big[  -\frac{1}{12} \left(4 \Sigma ^2+41\right) 
    \left(S^3+T^3+U^3\right) 
    \nonumber    \\&&\qquad
      +(\Sigma -1)   \left(m_s S^3+m_t T^3+m_u U^3\right)\Big]\zeta (5)+\cdots ~.\qquad
  \eeqn

 Furthermore, one   can verify that $\cA^{(1)}(S,T;m_s=p-1,m_t=0)$ agrees with $A^{(1)}$ computed from $\EV{pp11}$. This also gives a   consistency check. 

  The expression \eqref{cA1p1234} with coefficient \eqref{RRscof}\eqref{RRacof} is the main result of this paper. It gives the  worldsheet representation of the Virasoro-Shapiro amplitude of the four arbitrary KK modes in the $\AdS_3$ string theory supported by RR fluxes. 
  
Let us compare the $\AdS_3$ result here  with that of  $\AdS_5$.  In \cite{Wang:2025pjo}, it was shown that the final results can be rewritten in a  basis of weight-3 SVMPL which is free of singularities $\log^i 0$ for $i\le 3$. This  may be related to the maximal SUSY of $\AdS_5\times S^5$. This feature does not happen here: the 7 functions  $\cR_j^{s/a}$ are all linearly independent; this  may be related to the fact that   $\AdS_3$ solution is not maximally supersymmetric.

\section{Conclusion}\label{concl}

In this paper, we generalize the flat-space Virasoro–Shapiro amplitude to AdS space and derive the leading curvature correction to string amplitudes involving four arbitrary KK modes on $\AdS_3 \times S^3 \times M_4$ in type IIB string theory. The result takes the form of an integral over the Riemann sphere, analogous to the flat-space Virasoro–Shapiro amplitude, but with insertions of single-valued multiple polylogarithms. Our derivation is fully self-consistent and allows us to extract an infinite amount of CFT data in the D1-D5 CFT, which could be useful for future integrability and bootstrap studies.\footnote{The AdS VS amplitude with four arbitrary KK modes derived in Section~\ref{wsp1234} encodes even more CFT data, which can be extracted. We leave this analysis for future work. }
In particular, the scaling dimensions of certain operators predicted by our method are shown to be consistent with classical string theory computations under specific assumptions. 

In the case of $\mathcal{N}=4$ SYM, integrability—especially   the quantum spectral curve method—has provided a powerful and independent field-theoretic approach to computing scaling dimensions and OPE coefficients. Remarkably, these results have been shown to be consistent with the Virasoro–Shapiro amplitude in type IIB string theory on $\AdS_5 \times S^5$~\cite{Gromov:2023hzc}. It would therefore be highly significant to extend integrability techniques,\footnote{Integrability has been used to study the spectrum of the $T^4$ theory in the planar large $N$ limit.
For reviews of progress in this direction, see \cite{Sfondrini:2014via,Demulder:2023bux,Seibold:2024qkh}.
}
including the quantum spectral curve  method,\footnote{The quantum spectral curve for the $M_4 = T^4$ case was recently proposed in \cite{Cavaglia:2021eqr,Ekhammar:2021pys,Ekhammar:2024kzp}, but it is not yet developed to the stage where CFT data can be extracted at the strong coupling limit. 
Future advances in solving the quantum spectral curve  numerically, in the same spirit of the $AdS_5$ case \cite{Gromov:2023hzc,Julius:2023hre,Julius:2024ewf},   are expected to shed some new insight into the strong-coupling spectrum of $AdS_3 \times S^3 \times T^4$.
}
to study the D1-D5 CFT dual to string theory on $\AdS_3$, and compare with the results in this paper.

Given how little is currently known about the D1-D5 CFT, it would be interesting to generalize the worldsheet methods developed in this work further. One possible generalization  is to study additional half-BPS operators involving both the tensor and graviton multiplets.
It would also be valuable to understand curvature corrections beyond the leading order, or generalize the four-point amplitudes to higher points examined in this paper.
Another important, though challenging, direction is to extend the AdS Virasoro–Shapiro amplitude beyond tree level. In such a setting, the worldsheet is expected to be replaced by a torus or a higher-genus Riemann surface, and the single-valued multiple polylogarithms are supposed to be replaced by some modular analogues.  

Finally, it would be significant to uncover the worldsheet description of string theory on $\AdS_3$ supported by RR fluxes and to derive the Virasoro–Shapiro amplitudes directly from the worldsheet formulation. Alternatively, one may employ the formalism of string field theory to study AdS/CFT. Some  progress has been made for $AdS_3$, where the exact worldsheet formulation of string theory was proposed at pure NS-NS point, and contribution of RR flux can be computed perturbatively \cite{Cho:2018nfn}.\footnote{For recent developments in  string field theory  computations in the $AdS_5 \times S^5$ case, see \cite{Cho:2025coy}.}

\acknowledgments

We   thank Shai Chester, Tobias Hansen and  Arkady Tseytlin for discussions/comments on the draft.
HJ was supported by the startup grant at SIMIS, and  also  in part by the STFC
Consolidated Grants ST/T000791/1 and ST/X000575/1. DlZ is supported in part by the UK Engineering and Physical Sciences Research Council grant number EP/Z000106/1, and the Royal Society under the grant URF\textbackslash R1\textbackslash 221310.

\appendix

\section{Single-valued multiple polylogarithms} \label{apd-SVMPL}

The multiple polylogarithms (MPLs) of weight $r$ are defined though iterated integrals \cite{Goncharov:2001iea,Goncharov:1998kja},
\be
L_{a_1 a_2 \ldots a_r}(z) \coloneq \int_0^z \frac{\dd t}{t-a_1} L_{a_2 \ldots a_r}(t)~,
\ee
where the words $a_i$ take values in $\{0,1\}$. Explicitly, the multiple polylogarithms up to weight three are given in terms of combinations of logarithm, polylogarithms and zeta values as follows:
\be
\begin{aligned}
L_{0^p}(z) =&\frac{\log ^p z}{p!}~, \\
 L_{1^p}(z) =&\frac{\log ^p(1-z)}{p!}~, \\
L_{01}(z)= & -\operatorname{Li}_2(z)~, \\
 L_{10}(z)=&\operatorname{Li}_2(z)+\log (1-z) \log (z)~, \\
  L_{010}(z)=&2 \operatorname{Li}_3(z)-\operatorname{Li}_2(z) \log (z)~,\\
L_{001}(z)= & -\operatorname{Li}_3(z)~, \\
 L_{100}(z)= & -\operatorname{Li}_3(z)+\operatorname{Li}_2(z) \log (z)+\frac{1}{2} \log (1-z) \log ^2(z)~,\\
L_{101}(z)= & 2 \operatorname{Li}_3(1-z)-2 \zeta(3)-\log (1-z)\left(2 \operatorname{Li}_2(1-z)\right.   \left.+\operatorname{Li}_2(z)+\log (1-z) \log (z)\right) ~,\\
L_{110}(z)= & -\operatorname{Li}_3(1-z)+\frac{1}{6} \pi^2 \log (1-z)+\zeta(3)~, \\
L_{011}(z)= & -\operatorname{Li}_3(1-z)+\operatorname{Li}_2(1-z) \log (1-z)  +\frac{1}{2} \log (z) \log ^2(1-z)+\zeta(3)~,
\end{aligned}
\ee
where $\operatorname{Li}_n(z)$ denotes the polylogarithms of order $n$. 

MPLs are holomorphic functions of $z$ but they are not singe-valued. The single-valued version of MPLs can be constructed following the method in \cite{Brown:2004ugm}. For our purposes, it suffices to know that up to weight three, single-valued multiple polylogarithms (SVMPLs) $\mathcal{L}$ are constructed from MPLs as
\be
\begin{aligned}
& \mathcal{L}_{000}(z)=L_{000}(z)+L_{000}(\bar{z})+L_{00}(z) L_0(\bar{z})+L_0(z) L_{00}(\bar{z}), \\
& \mathcal{L}_{001}(z)=L_{001}(z)+L_{100}(\bar{z})+L_{00}(z) L_1(\bar{z})+L_0(z) L_{10}(\bar{z}), \\
& \mathcal{L}_{010}(z)=L_{010}(z)+L_{010}(\bar{z})+L_{01}(z) L_0(\bar{z})+L_0(z) L_{01}(\bar{z}), \\
& \mathcal{L}_{100}(z)=L_{100}(z)+L_{001}(\bar{z})+L_{10}(z) L_0(\bar{z})+L_1(z) L_{00}(\bar{z}), \\
& \mathcal{L}_{110}(z)=L_{110}(z)+L_{011}(\bar{z})+L_{11}(z) L_0(\bar{z})+L_1(z) L_{01}(\bar{z}), \\
& \mathcal{L}_{101}(z)=L_{101}(z)+L_{101}(\bar{z})+L_{10}(z) L_1(\bar{z})+L_1(z) L_{10}(\bar{z}), \\
& \mathcal{L}_{011}(z)=L_{011}(z)+L_{110}(\bar{z})+L_{01}(z) L_1(\bar{z})+L_0(z) L_{11}(\bar{z}), \\
& \mathcal{L}_{111}(z)=L_{111}(z)+L_{111}(\bar{z})+L_{11}(z) L_1(\bar{z})+L_1(z) L_{11}(\bar{z}) .
\end{aligned}
\ee

MPLs and SVMPLs can be implemented using package $\mathtt{PolyLogTools}$ \cite{Duhr:2019tlz,Maitre:2005uu}. The MPL is denoted as $\mathtt{G}$: $L_{a_1\cdots a_n}(z) \coloneq \mathtt{G}[a_1, \cdots, a_n,z]$. Similarly SVMPL is denoted as $\mathtt{cG}$: $\cL_{a_1\cdots a_n}= \mathtt{cG}[a_1, \cdots, a_n,z]$.

\section{Mack Polynomials} \label{apd-mackpol}

Following \cite{Dolan:2011dv}, see also \cite{Mack:2009mi,Fardelli:2023fyq,Dey:2017fab,Mack:2009mi}, the Mack polynomial $Q^{\Delta_{12}, \Delta_{34}, \tau}_{\ell, m}(s)$ admits the following double sum form:
\beqn
Q^{\Delta_{12}, \Delta_{34}, \tau}_{\ell, m}(s) &= 
 \sum\limits_{k=0}^\ell \sum\limits_{n=0}^{\ell - k}
(-m)_k \left( m + \frac{s + \tau}{2} \right)_n 
\mu(\ell, k, n, \Delta_{12}, \Delta_{34}, \tau)~,
\eeqn
where $(a)_b$ denotes the Pochhammer symbol, and $\mu$ is given by
\begin{align}
\mu(\ell, k, n, \Delta_{12}, \Delta_{34}, \tau) &=
(-1)^{k+n+\ell } 
\frac{2^{ \ell} \, \ell! \, \Gamma(\ell + \tau - 1)}{\Gamma(2\ell + \tau - 1)}
\frac{ (\ell + \tau - 1)_n}{k! \, n! \, (-k - n + \ell)!}
\left( n + \frac{d} {2}-1+ \frac{\Delta_{34} - \Delta_{12}}{2} \right)_k \notag \\
&\quad \times 
\left( k + n - \frac{\Delta_{12}}{2} + \frac{\tau}{2} \right)_{-k - n + \ell}
\left( k + n + \frac{\Delta_{34}}{2} + \frac{\tau}{2} \right)_{-k - n + \ell} \notag \\
&\quad \times 
{}_4F_3\left(
\begin{array}{c}
-k, \, 3-d - n - \ell, \, 1-\frac{d}{2} + \frac{\Delta_{12}}{2} + \frac{\tau}{2}, \, 1-\frac{d}{2} - \frac{\Delta_{34}}{2} + \frac{\tau}{2} \\
2-\frac{d}{2}-\ell, \,2-\frac{d}{2}+ \frac{\Delta_{12}}{2} - \frac{\Delta_{34}}{2} - k - n, \,  2-d + \tau
\end{array}
; 1
\right)\, .
\end{align}

 The normalization  of Mack polynomial is chosen such that $ Q_{\ell,m}^{\tau,d}(s)=s^\ell+\cdots$. For $m=0$, the Mack polynomial reduces to 
  \be
  Q_{\ell,0}^{\tau,d}(s)=  \frac{(-2)^\ell \left(\frac{\tau }{2}-\frac{\Delta_{12}}{2}\right)_\ell \left(\frac{\Delta_{34}}{2}+\frac{\tau }{2}\right)_\ell  }{(\ell+\tau -1)_\ell}\, _3F_2\left(-\ell,\frac{s}{2}+\frac{\tau }{2},\ell+\tau -1;\frac{\tau }{2}-\frac{\Delta_{12}}{2},\frac{\Delta_{34}}{2}+\frac{\tau }{2};1\right)~.\\
  \ee

The normalization factor $\kappa$ used in the main text reads 
\be\begin{aligned} 
\kappa_{\ell, m, \tau,d}^{\left(p_1, p_2, p_3, p_4\right)} & 
= \frac{-2^{1-\ell}(\ell+\tau-1)_{\ell} \Gamma(2 \ell+\tau)  }{m!(1-\frac{d}{2}+\ell+\tau)_m }
\frac{1}{ \Gamma\left(-m+\frac{p_1+p_2}{2} -\frac{\tau}{2}\right) \Gamma\left(-m+\frac{p_3+p_4}{2} -\frac{\tau}{2}\right)}
 \\ & \times 
 \frac{1}{ 
\Gamma\left(-\frac{p_1-p_2}{2}+\ell+\frac{\tau}{2}\right)
\Gamma\left(\frac{p_1-p_2}{2}+\ell+\frac{\tau}{2}\right)
 \Gamma\left(-\frac{p_3-p_4}{2}+\ell+\frac{\tau}{2}\right) \Gamma\left(\frac{p_3-p_4}{2}+\ell+\frac{\tau}{2}\right) 
}\, .
\end{aligned}
\ee
\section{Ambiguities of the worldsheet integrand}\label{zeromd}

We now list the five ambiguities in the worldsheet integrand \eqref{eqn-A1-GHansatz} that integrate to zero upon performing the worldsheet integration \eqref{eqn-A1-worldsheet}.
\begin{align*}
  1): & \quad
g^s = (0, 0, 0, S T), \\
& \quad g ^a = (0, 0, 0), \\
& \quad h^s = (0, 0, 0, 0), \\
& \quad h^a = (0, 0, 0); \\[1em]
 2): & \quad
g^s = (S^2 + 4 S T + T^2,\; -4(S^2 + S T + T^2),\; 4 S^2 - 2 S T + 4 T^2,\; -32(S^2 + T^2)), \\
& \quad g^a = ((S - T)(S + T),\; 4(S - T)(S + T),\; 0), \\
& \quad h^s = (0, 0, 0, 0), \\
& \quad h^a = (0, 0, 0); \\[1em]
 3): & \quad
g^s = (0, 0, 0, S^2 + T^2), \\
& \quad g^a = (0, 0, 0), \\
& \quad h^s = (0, 0, 0, -2 S^2 - T^2), \\
& \quad h^a = (0, 0, 0); \\[1em]
 4): & \quad
g^s = (S^2 + T^2,\; -4 S T,\; 2 S T,\; 0), \\
& \quad g^a = ((S - T)(S + T),\; 0,\; 0), \\
& \quad h^s = (2(S - T) T,\; 4(S - T) T,\; -4 S T + 4 T^2,\; 32 S T - 32 T^2), \\
& \quad h^a = (2 T(S + T),\; -4 T(S + T),\; 0); \\[1em]
 5): & \quad
g^s = (0,\; 0,\; S^2 - 2 S T + T^2,\; 0), \\
& \quad g^a = (0,\; 0,\; (S - T)(S + T)), \\
& \quad h^s = (S(S + 2 T),\; -2 T(S + 2 T),\; -2 S^2 + 2 S T + 3 T^2,\; -16 S T - 32 T^2), \\
& \quad h^a = (S(S + 2 T),\; -2 T(S + 2 T),\; -2 S^2 - 2 S T + T^2).
\end{align*}

\bibliographystyle{JHEP.bst} 
\bibliography{AdS3}

\end{document}